\documentstyle[preprint,aps,12pt]{revtex}

\begin{document}

\title{Strong Coupling Hadron Masses in $1/d$ Expansion for Wilson fermions.}

\author{B.~Rosenstein}
\address{Department of Physics, National Tsing - Hua University, Hsinchu,
Taiwan, R.O.C.}

\author{ A.~D.~Speliotopoulos}
\address{Institute of Physics, Academia Sinica\\ Taipei, Taiwan 115, R.O.C.}
\maketitle
\begin{abstract}
Motivated by the weak-strong coupling expansion
\cite{Rosenstein}, we calculate the spectrum of hadrons using a
systematic $1/d$ ($d$ - dimensionality of spacetime) in addition to
a strong coupling expansion in $\beta$. The $1/d$ expansion is
pushed to the next to leading order in ($1/d$) for mesons
and next to next to leading  order for baryons. We do the calculation
using Wilson fermions with arbitrary $r$ and show that doublers decouple
from the spectrum only when $r$ is close to the Wilson's value $r=1$.
For these $r$ the spectrum is much closer to the lattice results
and the phenomenological values than those obtained by using either
the (nonsystematic) "randomwalk" approximation or the hopping
parameter expansion. In particular, the
value of the nucleon to $\rho$ - meson mass ratio
 is lowered to $\frac {3 \log d -1/4}{2 {\rm arccosh} 2}+O(1/d)\approx 1.48$.
The result holds even for $\beta$
as large as 5, where the weak-strong coupling expansion is
applicable and therefore these results are expected to be reasonable.
\end{abstract}

It is commonly believed that for low energy physical
quantities in QCD, such as hadron masses, there is no small expansion
parameter, and the theory is "strongly coupled".
The theory only has an asymptotic weak coupling
expansion successfully describes high energy
quantities but breaks down at low energies. Recently one of us
proposed a scheme combining weak and strong coupling
expansion in a "double expansion scheme" \cite{Rosenstein}.
The high energy modes are integrated out using
expansion in weak coupling $\alpha_s$,
the resultant effective Lagrangian is then expanded in derivatives
and solved using strong coupling expansion in $\beta$.

{\it A priori} it would seem that the domain of applicability for the
expansion in $\beta$ will have very little chance to intersect with
that of the expansion in $\alpha$. Indeed, if the gauge coupling constant
$g$ is small,  $1/g$ is large and {\it vice versa}. Fortunately,
however, a small $\alpha(g)$, does not necessarily imply that
$\beta(g)$ will be large and {\it vice versa}. In fact, this scheme of
a simultaneous weak and strong expansion has been tested on certain
solvable low dimensional asymptotically free models like Ising chain
and $d=2$ Gross - Neveu models \cite{Rosenstein} and the results agree
very well with the exact values.

As for QCD, looking at strong coupling expansion results \cite{Creutz,Kogut}
one notices that although the asymptotic scaling region is out of reach, the
effective weak coupling $\alpha$ near the strong coupling radius
of convergence is quite small. In the $SU(3)$ YM theory the radius of
convergence (roughening phase transition), is larger than $\beta_{max}\equiv
6/g_{min}^2\sim 5$ \cite{Munster} which corresponds (using naive
perturbative RG) to a relatively small effective weak coupling
expansion parameter $\alpha_{lat}\equiv g_{lat}^2/4\pi\sim 0.1$. Even
taking into account the fact that $\alpha_{lat}$ is a lattice one
(not the $\overline{\rm MS}, \alpha_{\overline{\rm MS}}\sim 0.3$),
this corresponds to the values at which perturbation theory is
supposed to work for energies above $1.5-2$ GeV. Recently, this fact
has been fully understood for lattice weak coupling perturbation
theory \cite{MKL}. Therefore, there exists a (albeit smaller then the
one for the Gross - Neveu model) window in the coupling in
which both $\alpha$ and $\beta$ are small enough to produce a
reasonable series. The well known ``loop factors'' $1/(4\pi)^2$ in the
weak coupling expansion parameter $\alpha$ are partly responsible for
this, although this generally is not sufficient since symmetry group
factors tend to reduce the window. Note also that the leading order
term in $\alpha$ coincides with  the conventional ``phenomenological''
strong coupling model, in which the inverse lattice spacing $M$ is
limited to values inside the weak-strong applicability window. This
would seem to be a strong indication that the simultationeous
weak-strong expansion will be applicable to QCD.

The complexity of such a calculation depends on the quantity
and the precision one would like to achieve. Usually strong coupling
series are relatively easy to evaluate to very high orders. Consequently
the scale $M$ can be chosen in such a way that $\beta_M$ is just below
the strong coupling expansion radius of
convergence. Simultaneously, the scale $M$ should
be sufficiently high or alternatively the relevant energy scale
sufficiently low so that just a few orders in the derivative expansion are
needed to achieve the desired accuracy. Consequently, this method is
limited to low energy quantities only. Therefore, inside this
weak-strong coupling window the usual lattice action with renormalized
coupling can serve as a reasonable effective action \cite{Wilson}.
The expected accuracy is rather low: up to corrections
of order $\alpha_{\overline {\rm MS}}\sim 0.3$ due to weak coupling
expansion. On the top of this we moreover will then have to perform
the strong coupling expansion. Fortunately it is well known that in
the hadronic sector (unlike the glueball or pure glue sector in which
the above estimates of the radius of convergence of the strong
coupling were taken) the situation is much better. The strong coupling
limit $\beta=0$ already produces a reasonable spectrum of hadrons. In
addition, it is known that the next to leading order terms in $\beta$
for the hadronic spectrum are numerically very small \cite{HS,KS}.
Even for the window value of $\beta=5$ the corrections do not exceed
15\%.

However even in this limit QCD is nontrivial and an additional expansion
parameter should be utilized. This may be the hopping parameter,
$1/N_c$ or $1/d$ expansions. The hopping expansion parameter
$\kappa\approx 1/2$ is defined as a bare coupling in units of lattice
spacing and therefore cannot be easily related to a physical quantity
\cite{Creutz,Hasenfratz}. The $1/N_c=1/3$ has been extensively used,
but is notoriously  difficult to perform beyond the leading order.
Indeed, it is mostly the nonsystematic random walk approximation
\cite{KS,HS,G} that has been used to estimate the spectrum at
strong coupling. In this approximation the hopping parameter expansion is
partially summed up, so that quark-lines form ``collapsed paths''
\cite{G,HS}. Although it seems to be superior than the
simple hopping parameter expansion, one does not find a controllable
expansion parameter within this approximation. Moreover, the results of this
approximation as well as those of the hopping parameter expansion were
rather discouraging. Although the ordering of the lowest hadronic
states is correct, some mass ratios are grotesque. An especially bad
example is the nucleon to $\rho$ - meson mass ratio which is about 2.2
instead of the phenomenological 1.2 or (at $\beta=5$) lattice MC
simulation value of $\sim 1.4$.

In this paper we use the $1/d=1/4$ expansion to calculate the hadron mass
spectrum, which allows managable higher order calculations. This was
first applied to Yang-Mills theories for
staggered fermions in \cite{Stern}. The results for the nucleon to
$\rho$ mass ratio is $\sim 1.7$, which, although better the
previously mentioned strong coupling result is still so different
from the phenomenological value that it cannot be accounted for by
next to leading order corrections. Moreover, the interpretation of
particle spectrum for staggered fermions is by no means straightforward.
We, therefore, shall consider Wilson fermions which
allows us to discuss splitting due to three flavors and for
which the interpretation is straightforward.

In view of these above results for the spectrum, one of the following
should be valid: (a) something is nevertheless wrong with the argument
for the existence of the weak-strong coupling window {\it and\/} the
quenched lattice results are not precise enough. The spectrum at
$\beta=5$ is indeed very different from the observed experimental one
because the continuum limit is still far from this point or
(b) the random walk and the hopping expansion (and to a smaller degree
the staggered fermion) results are inaccurate.
We show in this paper that when a systematic $1/d$ expansion is applied
to the Wilson action within the weak-strong window, a spectrum is
obtained which is in agreement with the above lattice MC results
within the expected accuracy of the expansion.

We now fix notation and outline the formalism, which is well-known,
focusing on the differences with random walk approach.
No details of higher order calculations will be given.
The standard lattice Wilson action is
\begin{equation}
{\cal S} = -\frac{\beta}{2N_c}\sum_{plaquettes} \left(Tr U_{\Box} + Tr
U^\dagger_{\Box} \right) + \sum_{x,\mu}
\left\{\overline J^{\cal AB}_\mu U^{\cal BA}_\mu + {U_\mu^\dagger}^{\cal
AB} J^{\cal BA}_\mu\right\}
-\sum_x m \overline\psi^{{\cal A}}_a \psi ^{{\cal A}}_a
\label{e1}
\end{equation}
where $J^{\cal AB}_\mu(x) \equiv \overline\psi^{\cal A}_a(x+\mu)
P^{+}_\mu\psi^{\cal
B}_a(x)$ and $P_\mu^{\pm} = (r\pm\gamma_\mu)/2$. $U^{\cal AB}$ is the usual
compact gauge field on the lattice and the script letters run over
$N_c$ colors. The lower latin letters runs over $N_f$ flavors.
First, we shall limit ourselves to the $\beta=0$ limit
and then discuss the effects of finite $\beta$.

Integration over the gauge fields $U$ give in the leading order
$\cite{KS}$,
\begin{equation}
{\cal S}_0 = -\frac{1}{N_c}\sum_{x} \sum_{\mu=1}^d\overline J^{\cal
AB}_\mu J^{\cal BA}_\mu (x)- \frac{1}{N_c!}\sum_{x,\mu}
\left[\det_c{J_\mu^{\cal AB}}(x) + \det_c{\overline J_\mu^{\cal AB}}(x)\right]
+\dots
\label{e3}
\end{equation}
where the determinant is over the color indices. The $\dots$
indicates a finite number of terms which contribute to higher order
terms in $1/d$. Considering first
the mesonic sector, we note that although we have introduced different
flavors, for the
leading order in $1/d$, they will not play a role and we shall
suppress them for now. We then introduce the mesonic fields through
$M^A(x) = \frac{1}{N_c}\sqrt{\frac{d}{2}}\>\>\overline\psi^{\cal
A}\Gamma^A\psi^{\cal A}$.
The ``channel'' index $A$ runs from $0$ to $15$ and normalization of matrices
$\Gamma$
is chosen so that ${\rm Tr}\>\Gamma^A\Gamma^B =
\delta^{AB}$ \cite{HS}. In terms of these fields the mesonic part of the action
becomes
\begin{equation}
{\cal S}_0^{mes} = -\frac{N_c}{2}\sum_{xy\mu} M^A(x) D^{AB}(x-y)
M^B(y)
\label{e8}
\end{equation}
where unless otherwised stated summation over repeated channel
indices will be understood.
We now integrate over the fermion fields. This is done by first
introducing auxillary fields conjugate to $M^A$ \cite{com5}:
\begin{equation}
e^{-{\cal S}_0^{mes}} = \int {\cal D M}^A
\exp\left\{N_c\sum_{x}M^A(x){\cal M}^A(x)
- \frac{N_c}{2} \sum_{xy}{\cal M}^A(x) D^{-1}_{AB}(x-y){\cal
M}^B(y) \right\}
\label{e11}
\end{equation}
Then the remaining gaussian integral over fermionic fields can be
done:
\begin{eqnarray}
{\cal Z}_0 &=& \int {\cal D M}^A \exp\left\{- \frac{N_c}{2}
\sum_{xy} {\cal M}^A(x) D^{-1}_{AB}(x-y){\cal
M}^B(y)-N_c\sum_x{\rm Tr}_D\log\left(\Gamma^A {\cal M}^A(x) +
2\overline m \right)\right\}
\nonumber \\
&{}&\equiv  \int {\cal D M}^Ae^{-{\cal A}[{\cal M}(x)]}
\label{e12}
\end{eqnarray}
where $\overline m = m/\sqrt{2d}$. The factor $\sqrt d$ was
introduced in the mass to facilitate the $1/d$ expansion \cite{Stern}.
The functional ${\cal A}[{\cal M}(x)]$
of hadronic fields is an effective hadronic action describing dynamics
of the color invariant "basic" fields only. In the meson sector, which
is being considered now, these  fields interpolate between the
pseudoscalar and vector mesons. They correspond to the lowest energy
states of the naive quark model. Later on baryonic fields
interpolating between the octet and decouplet ($N$ and $\Delta$) fields will be
introduced.

The quadratic part of the mesonic action to lowest order is then
\begin{equation}
{\cal A}_0^{mes} = - \frac{N_c}{2} \sum_{xy}{\cal M}^A(x)
G^{-1}_{AB}(x-y){\cal M}^B(y)
\label{e15}
\end{equation}
where
\begin{equation}
G^{-1}_{AB}(x-y) = D^{-1}_{AB}(x-y) + \frac{\delta_{AB}}{\lambda_0^2}
\label{e16}
\end{equation}
and $\lambda_0=\overline m +\sqrt{\overline m +
1-r^2}$ comes from the solution of the gap equation \cite{Stern}.
To find the mass spectrum of the theory, we need to find the zeros of
$G^{-1}$ in momentum space. This would seem to be difficult, but note
that $G^{-1}$ and $k\equiv \lambda_0^2 G^{-1}D$ have the same zero
eigenvalues, as long as $D$ does not vanish and finding the zeros of
the latter quantity is fairly straightforward.

There are two coupled channels for the mesonic sector. First, the
pseudoscalar couples with the time component of the axial vector
giving the mass term for the pion. In this channel,
\begin{equation}
G^{-1}D\lambda_0^2 = \left(\matrix{
	\lambda_0^2-(1+r^2)+\frac{(1+r^2)}{d}\left(1-\cosh
	m\right)& -i\frac{2r}{d}\sinh m\cr
	-i\frac{2r}{d}\sinh m&\lambda_0^2+(1-r^2)-\frac{1-r^2}{d}-
	\frac{(1+r^2)}{d}\cosh m\cr}
	\right)
\label{e17}
\end{equation}
where we included the correct powers of $1/d$ and have taken the $d$-
momentum to be $(0,...,0,im)$. Within the $1/d$ expansion, we can generically
write
\begin{equation}
\cosh[m] = xd+y+O(1\d).
\label{cosh}
\end{equation}
where, of course, $y$ cannot be determined as yet since higher orders
in $1/d$ terms have not been included. Using eq.($\ref{cosh}$) and
expanding the eigenvalue equation of the matrix eq.(7) in orders of
$1/d$, we find two solutions. The one which is finite for all relevant
$r$, is
\begin{equation}
x_\pi = \frac{1}{(1-r^2)^2}
	\left\{
		(\lambda_0^2-r^2)(1+r^2)-
		\left[
		(1-r^2)^2+4r^2(\lambda_0^2-r^2)^2
		\right]^{1/2}
	\right\}
\label{e18}
\end{equation}
and determines the mass of the mass of the pion. The second eigenvalue
$x_d$ is nonzero and describes a doubler (it is a bound state of two
fermionic doublers at the opposite corners of the Brillouin zone):
\begin{equation}
\cosh m_d = \frac {2 d (1+r^2)}{(1-r^2)^2}
\label{e26}
\end{equation}
 One then requires
that the pion be massless for all $r$, and sets $x_\pi=0$ and
$y_\pi=1$. This, to lowest order,
determines the trajectory in parameter space relating
bare mass to $r$:
$\lambda_0^2 = 1+r^2$.

In the $\rho$ meson channel the corresponding matrix
(on the trajectory) is:
\begin{equation}
 \left(\matrix{\lambda_0^2-(1+r^2)+\frac{1}{d}\left(3+r^2-(1+r^2)\cosh
	m\right)& -i\frac{2r}{d}\sinh m\cr
	i\frac{2r}{d}\sinh m&\lambda_0^2 (1-r^2)+\frac{1}{d}
[-3+r^2-(1+r^2)\cosh m]\cr}
	\right)
\label{K}
\end{equation}
where we have considered only one spatial component of the
vector field interpolating $\rho$ mesons.
On this trajectory
we write $\cosh m_\rho=x_\rho d + y_\rho +O(1/d)$ and obtain
\begin{equation}
 \left(\matrix{-(1+r^2)x_\rho& -2i r x_\rho
	\cr 2i r x_\rho&2-(1+r^2) x_\rho\cr}
\right)\label{e25}
\end{equation}
to the leading order in $1/d$. There are two eigenvalues to this
matrix. The first is $x_\rho=0$. This means that the  $\rho$ meson's
mass does not have a ``natural'' order of $\log d$, but is , in fact,
smaller -- just
of order  1 . It is, however, inconsistent to to  determine $y_\rho$
by solving equation for
vanishing determinant of eq.(\ref{K}). The $1/d$ corrections to this
matrix, considered in the following, must be taken into account. Other
channels, scalar, tensor, etc, contain doublers only.

Genericly, in any next to leading order calculation in $1/d$ there are
two types of contributions \cite{Stern}. The first is the
"tree" contribution which arises from additional terms in the integral
over gauge fields eq.(\ref{e3}), while the second is the one loop diagrams
involving the propagator and vertices
of the leading effective action eq.(\ref{e12}).
Keeping the pion mass zero, the ``next to leading'' contribution to
the $\rho$ meson's mass $y_\rho$ (which is actually the leading since
$x_\rho=0$) is:
\begin{equation}
y_\rho=\frac{3+r^2}{1+r^2}\label{e17}
\end{equation}
This value is consistent with the $r=0$ result obtained in \cite{Stern}
for staggered fermions. Note that for $r=1$ the $\rho$ mass is
significantly larger then in the random walk approximation.

The purpose of introducing the chiral symmetry breaking mass and
Wilson's terms was to remove doublers. The value of $r$ should, in
principle, be optimized in such a way that on the one hand doublers do
not interfere with physical particles and, on the other hand,
the chiral symmetry is minimally violated. As is well known,
setting the pion mass to zero does not mean that the chiral
 symmetry is somehow restored on the trajectory. It just means  that we
are situated on the spontaneous parity breaking phase transition line.
On Fig.1
we show the $r$ dependence of various doublers masses compared to the
$\rho$ meson mass.
We can see that the doublers are significantly
 heavier then $\rho$ mesons only near Wilson's value
of $r=1$, where they
all become infinitely heavy.
Therefore we conclude that there is no great
advantage to work with
$r<1$ contrary to some lattice \cite{Weingarten} and random
 walk approximation \cite{G} results in which doublers
 were not considered.

Turning our attention to the baryonic sector, we note that the general
expression for the exponent of the baryon masses contains, half
integer as well as integer powers of $1/d$:
\begin{equation}
e^m=x d^{3/2}+x' d+yd^{1/2}+y' + O(1/\sqrt d)
\end{equation}
 We calculated $x,x', y$
and $y'$.
To the leading order
 the $\Delta$ mass
 on the $m_\pi=0$ trajectory is
\begin{equation}
 x_\Delta=\frac{2\sqrt{2}(1+r^2)^{3/2}}{(1+r)^3}
\label{delta}
\end{equation}
Notice that for $r=0$, $x_\Delta=2\sqrt 2$ which reproduces
the result obtained in \cite{Stern} for staggered fermions calculation.
The nucleon is degenerate with $\Delta$ up to the order of $1/d$ we
have considered  when $r=1$. The formulas for the higher order
corrections to the baryon masses for other values of $r$ are
cumbersome and will be given elswhere. Instead, for $r=1$
\begin{equation}
x_\Delta'=-\frac{1}{4},\   \  y_\Delta=-\frac {1}{12}, \   \
y_\Delta'=\frac {29}{144}
\label{masses}
\end{equation}
We also calculated the leading order $\beta$ corrections to the $\rho$
and baryons. These types of corrections has been studied for staggered
fermions in \cite{Stern} and within the random walk approximation
scheme in \cite{HS}. They vanish for $r=1$ and are small for other
values of $r$ near the window range.

To summarize, we have systematically studied the spectrum of the
Wilson action with arbitrary $r$ using strong coupling and the $1/d$
expansions. The Lagrangian is considered as a phenomenological
low energy effective Lagrangian for the values of $\beta$
at which the expansion is still reasonable. It turns out that the doublers
decouple from the physical spectrum only when $r$ is quite close to
the Wilson's value $r=1$. The results of the systematic $1/d$
expansion for the nucleon to $\rho$ mass ratio is $\frac {3 \log d
-1/4}{2 {\rm arccosh} 2}+O(1/d)\approx 1.48$. This value is within the
range of the next correction of the Monte Carlo results.

We now compare the $1/d$ results for mesons with those obtained within the
random walk approximation in $d=4$. To understand the difference with
the random walk approach, we have extended the random walk calculation
\cite{HS} to arbitrary $d$ and obtained the mass of $\rho$ meson for
$r=1$ as:
\begin{eqnarray}
\cosh m_\rho =
\frac{2d-1}{d+1}=2-\frac{3}{d}+\frac{3}{d^2}-\frac{3}{d^3}+\dots
\label{c1}
\end{eqnarray}
The first term in this expansion coincides with the systematic $1/d$
expansion. There is however no reason to expect that the
next order term in $1/d$ will resemble that in eq.(\ref{c1}).
{\it A priori\/} this is not obvious since at the special value $r=1$,
many contributions to the next to leading order term vanish. This is
due to the well-known result from the random walk approximation
\cite{HS} that to this order all the propagators
travel along a single direction. As such, we will necessarily
encounter product of two projectors $P^+_\mu P^-_\mu=0$.
However if we go to the next to next to leading order there will
certainly be many contributions which will not vanish for the simple
reason that certain diagrams will {\it not ``linear'', but rather
``planar''}.

The coefficient in front of $1/d$ in eq.(\ref{c1}) is negative and
large numerically for $d=4$. This leads to significant underestimate
for the $\rho$ mass and consequently for the overestimation of
$m_N/m_\rho$. Let us emphasize that this term cannot be taken
seriously at this point since corrections to the mass at this order in
$1/d$ have not been done as yet. Indeed, it would be very
interesting to calculate this next ( $1/d$ for $\cosh m_\rho$)
order term to obtain better estimate of the $\rho$ mass.

This feature of random walk approximation does not carry over
to baryon sector, however. The corresponding expansion for the
nucleon mass in powers of $1/d$ is:
\begin{eqnarray}
e^{m_N}=d^{3/2}-\frac{1}{4}d-\frac{5}{48}d^{1/2}+\frac{119}{576}+\dots
\label{c2}
\end{eqnarray}
The first {\it two} terms of these now coincide with our systematic
$1/d$ expansion results. When a similar expansion is done for the
Delta mass, we find that for some unknown reason all {\it four\/}
terms now coincide with eq.(\ref{masses}). Once again, however, any
terms which is of higher order than $d$ in eq.~(\ref{c2}) are
unreliable, since they will be almost certainly be changed by higher
order $1/d$ corrections. In particular, we see that small splitting
between the nucleon and $\Delta$ found in random walk approach is due
these unreliable higher order terms. Within the systematic $1/d$
expansion, the Delta and Nucleon are degenerate.

Of course there are numerous other corrections to the weak-strong
approximation scheme. Here we discuss few of the many which have not
been calculated. Our previous discussion revealed the peculiar fact
most of the contributions to the next to leading order terms in $1/d$
vanish for $r=1$. We can ask the following question: What are the
corrections that do not vanish? In particular, we note that to the
next to leading order there is no splitting between Goldstone bosons
and flavour singlet - the $\eta$ particles. As is well known, this
splitting is due to an anomaly. The plaquette correction also does not
lead to the splitting. This is easy to understand. The arguments of
Frohlich and King \cite{FK} are applicable to the $1/d$ expansion. We
therefore expect these anomaly effects to appear only at very high
orders in $1/d$ or $\beta$, when a diagram which spreads in all four
directions can be constructed. Another possibility is that the
main mechanism for the splitting is not due to these corrections but
rather to direct anomaly breaking terms which are proportional to
$\alpha_s$. These appear due to the presence of instantons at energies
higher then the scale $M$.

The actual splitting between $\Delta$ and the nucleon is also probably
due to higher order terms in the effective Lagrangian. Note that in lattice
simulations at relatively small $\beta$ ($\beta\sim 5$) the splitting
is also smaller then the phenomenological values. The presence of
higher dimension terms are also crucial for two other purposes:
restoration of the Lorentz invariance and the chiral symmetry. Chiral
symmetry is explicitly broken by the mass and Wilson terms. As we have
mentioned, the vanishing of the pion mass is not sufficient tp restore
chiral symmetry. Instead it is simply a signal criticality with
respect to a discrete symmetry. It would be therefore be interesting
to investigate whether the chiral properties are gradually
restored once higher dimensional operators are introduced. For example
the pion scattering at small momenta is nonvanishing \cite{KS} without
higher dimensional terms. It is reasonable to expect that with the
inclusion of the term due to next order correction in $1/M^2$ of the
derivitive expansion of the effective action the correct zero
momentum limit at least will be recovered. Existing results, which are
quite scarce within the random walk approximation scheme, in which
only part of the higher dimensional (improvement) operators are
considered did not address this question.

\begin{center}
{\bf Acknowledgements}
\end{center}

We are indebted to A. Kovner, T. Bhattacharrya, L. Lin and H.-L. Yu for
numerous discussions. B. R. was supported by the NSC of ROC, grant
NSC-83-0208-M-001-011 while A.D.S was support by NSC Grant No.
NSC-84-2112-M-001-022.

\pagebreak

\vspace{18 cm}
Fig.1 Masses of nucleon, $\rho$ meson, mesonic and baryonic doublers
as function of Wilson parameter $r$.

\end{document}